\def\ba{\begin{eqnarray}}
\def\ea{\end{eqnarray}}
\def\n{\nonumber \\}
\documentclass[12pt]{article}
\usepackage{epsfig}

\begin{document} 
\title{ Moments of Wigner function
 and Renyi entropies   at freeze-out}

\author{A.Bialas, W.Czyz and K. Zalewski \\ M.Smoluchowski 
Institute of Physics \\Jagellonian
University, Cracow\thanks{Address: Reymonta 4, 30-059 Krakow, Poland;
e-mail:bialas@th.if.uj.edu.pl;}
\\ Institute of Nuclear Physics, Polish Academy of Sciences
\thanks{Radzikowskiego 152, Krakow, Poland}}
\maketitle

\begin{abstract}

Relation between  Renyi entropies and moments of the Wigner function,
representing the quantum mechanical description of the M-particle
semi-inclusive distribution at freeze-out, is investigated. It is shown
that in the limit of infinite volume of the system, the classical and
quantum descriptions are equivalent. Finite volume corrections are
derived and shown to be small for systems encountered in relativistic
heavy ion collisions.

\end{abstract}

\section{ Introduction}
 
It is widely recognized that  information on the entropy of the systems
produced in high-energy collisions is very useful in understanding 
the physics of the process in question. This is particularly important
for heavy ion collisions and  search for quark-gluon plasma. It was
suggested some time ago \cite{bc} that measurements of coincidences
between the events observed in high-energy experiments may provide an
estimate of the Renyi entropies \cite{ren} of the final state and thus
 also access information on its thermodynamical (Shannon) entropy.
Although the idea seems attractive, the argument of \cite{bc}, being
essentially of classical nature, can only be considered as a first step.
The proper formulation must take into account the quantum-mechanical
nature of the problem. In the present paper  a quantum-mechanical
formulation is developed and its consequences discussed. In particular,
it is shown that the classical approach of \cite{bc} is valid in the
limit of very large size of the system in configuration space. For
systems of finite size, quantum corrections are derived and shown to be
relatively small for final states occuring in relativistic heavy ion
collisions.

The order $l$ Renyi entropy, $H(l)$, of a  statistical system
 is defined as 
\ba H(l)= \frac 1{1-l}\log C(l), \label{i1}
\ea 
where  $C(l)$ are 
coincidence probabilities of the states of the system, given by
\ba
C(l)\equiv \sum_i[P_i]^l= Tr [\rho]^l. \label{i2}
\ea
The sum runs over all states of the system, $P_i$ is the
probability of a state $i$ to occur and $\rho$ is the density matrix
of the system\footnote{The second part of this equality is best seen in
the representation where the density matrix is diagonal.}.

The attractive property of Renyi entropies is their relation to the 
standard (Shannon) entropy of the statistical system. It is easy to show 
that
\ba
S\equiv Tr [\rho \log \rho ] = \lim_{l\rightarrow 1} H(l) ,          \label{i3}
\ea
where $S$ is the Shannon entropy. 

Moreover since, as is well known \cite{beck}, for $l\geq 1$  
\ba S \geq H(l)\geq H(l+1),    \label{i4} 
\ea 
the Renyi entropies provide an {\it exact} lower limit for $S$, a quantity
very important for understanding the properties of the quark-gluon
plasma \cite{plasma}.

The object of our investigation is an M-particle statistical system,
i.e., a collection of M-particle final states which we define as those
in which exactly $M$ particles were observed in a given region of the
momentum space. We shall call them $M$-particle events (independently of
how many particles were actually produced)\footnote{ This terminology is
often used in experimental descriptions of multiparticle processes. For
the momentum distributions, the proper technical terms are: exclusive
distribution if all the particles are observed, and semi-inclusive
distribution if besides a given number of observed particles there is an
unspecified number of other particles. The latter should not be confused
with inclusive $M$-particle distributions.}.

On the classical level, M-particle final states can be described by the
normalized $M$ particle phase-space distribution $W^{class}(X,K)$ with
$X=X_1,...,Z_M$, $K=K^{(1)}_x,...,K^{(M)}_z$.\footnote{Even at the
classical level, however,  the phase-space distribution of particles
produced in high-energy scattering is not a precisely defined quantity:
one has to take into account that particles may be produced at different
times. In the present paper, following \cite{ber1,bpb}, we are
considering the time-averaged distribution.}


To obtain a quantum mechanical generalization of the phase-space
distribution we follow the standard procedure, where the proper
quantum description of an M-particle final state is given by the
density matrix $\rho(p_1,...,p_M;p_1',...,p_M')$. As shown by Wigner
\cite{wig} the quantum-mechanical analogue of the classical phase-space
distribution (called Wigner function), can be defined in terms of
the density matrix as the Fourier transform
\ba
W (X,K) = \frac1{(2\pi)^{3M}} \int dq e^{-i qX} \rho(p,p') ,  \label{e2}
\ea
where $K=(p+p')/2;\; q=p-p'$. The quantum-mechanical description
of multiparticle events is obtained by considering the Wigner function
$W(X,K)$ instead of the classical phase-space distribution
$W^{class}(X,K)$.

The main goal of  this paper is to discuss the relation between 
the coincidence probabilities $C(l)$ defined in (\ref{i2}) and
 the moments of  the Wigner function
\ba
\hat{C}(l)= (2\pi)^{3M(l-1)} \int d^{3M}X \int d^{3M}K [W(X,K)]^l .  \label{e1}
\ea
Such a relation exists because both $C(l)$ and $\hat{C}(l)$ are defined
in terms of the density matrix. 

The factor $(2\pi)^{3M(l-1)}$ is introduced in (\ref{e1}) to account for
the factor $(2\pi)^{-3M}$ in (\ref{e2}) (which insures consistency
between the proper normalization of the density matrix and of the Wigner
function)\footnote{Actually, these factors involve $2\pi \hbar$. As is
customary, we have put $\hbar=1$.}. This choice guarantees that, when
$W(X,K)$ in (\ref{e1}) is replaced by the classical density distribution
$W^{class}(X,K)$, $\hat{C}(l)$ reproduces the correct classical limit of
the coincidence probability, thus satisfying an important consistency
constraint. We have
\ba 
\hat{C}(l)\rightarrow C^{class}(l)\equiv \int d^{3M}X  d^{3M}K
W^{class}(X,K)\left[(2\pi)^{3M}W^{class}(X,K)\right]^{l-1}. \label{ecl} 
\ea 
In this formula $W^{class}(X,K)$ can be interpreted as probability
distribution and $(2\pi)^{3M} W^{class}(X,K)$ as the
probability for the system to occupy the elementary phase-space cell
$(2\pi\hbar)^{3M}$ around the (6M-dimensional) point $(X,K)$. Thus one
sees that, indeed, (\ref{ecl}) represents the classical formula for the
coincidence probability.

The interest in the relation between $\hat{C}(l)$ and $C(l)$ stems from
the observation that the moments $\hat{C}(l)$ are experimentally
accessible: we have shown recently \cite{bck} that, for a rather large
class of models, $\hat{C}(l)$ defined above can be estimated from the
measured coincidence probability $C^{exp}(l)$ of the events with $M$
particles \cite{bc,dl,bc2}
\ba 
C^{exp}(l) =\frac {N_l}{N(N-1)...(N-l+1)/l!}, \label{31} 
\ea 
where $N_l$ is the number of the observed l-plets of identical events
and $N$ is the total number of events. $N(N-1)...(N-l+1)/l!$ is the
total number of l-plets of events\footnote{For l=2 formula (\ref{31})
was first suggested, in a different context, by Ma \cite{ma}. See also
\cite{bcw}.}. One sees that measurement of $C^{exp}(l)$ reduces to
counting the coincidence between observed events.

We show in this paper that $\hat{C}(l)$ (which can be measured by
counting the number of identical events \cite{bck}), and $C(l)$ (which
determine Renyi entropies and thus open a window to the true entropy)
are equal in the limit of infinite size of the system. Also the finite
volume (quantum) corrections are studied and shown to fall with inverse
square of the smallest (linear) size.

In the next section a model for the Wigner function is introduced and the
corresponding formulae for the moments $\hat{C}(l)$ are written down. In
Section 3 the results of \cite{bck} are summarized. In Section 4 the
coincidence probabilities $C(l)$ are analyzed in the same framework. The
relation between $\hat{C}(l)$ and $C(l)$ are discussed in Section 5. Our
conclusions and outlook are given in the last section.

\section{Moments of the Wigner function}

In terms of the Wigner function $W(K,X)$,
 the momentum distribution is given by the integral
\ba
w(K)\equiv e^{-v(K)}= \int d^{3M}X W(X,K).   \label{6c}
\ea

To discuss $\hat{C}(l)$, we follow the argument of \cite{bck} and
consider the time-averaged Wigner function of the rather general form,
valid in a large variety of models \cite{mod}:
\ba
W(X,K)= \frac1{(L_xL_yL_z)^M}G[X/L] w(K)           \label{36a}
\ea
with $X/L\equiv (X_1-\bar{X}_1)/L_x,...,(Z_M-\bar{Z}_M)/L_z$,
 $K= K_1,...,K_{M}$. The function $G$ 
satisfies  the normalization conditions
\ba
\int  G(u) d^{3M}u&=& 1\;\;\rightarrow\;\;\int dX G(X/L) =
(L_xL_yL_z)^M ;\n
\int  u_i G(u)d^{3M}u&=& 0\;\;\rightarrow\;\;
<X_i,Y_i,Z_i>= \bar{X}_i,\bar{Y}_i,\bar{Z}_i;\n
\int  (u_i)^2 G(u)d^{3M}u&=& 1\;\;\rightarrow\;\;
<(X_i-\bar{X}_i)^2,...> =L_x^2,...   \label{37}
\ea

The first condition insures that $w(K)$ is the correctly normalized
(multidimensional) momentum distribution, the second defines the central
values of the particle distribution in configuration space and the third
defines $L_x,L_y,L_z$ as root mean square sizes of the
distribution in configuration space. Both sizes and central positions
may depend on the particle momenta\footnote{They may also be different
for different kinds of particles.}. The form of function $G$ is
responsible for the shape of the multiparticle distribution in
configuration space\footnote{The distribution in configuration space is
given by $\int dK W(X,K)$ \cite{wig} and thus it is not {\it identical}
to {G(X/L)}.}.

Using (\ref{36a}) we obtain from (\ref{e1}) 
\ba
\hat{C}(l)=(2\pi)^{3M(l-1)} \int d^{3M}K [w(K_1,...,K_M)]^l 
\int\frac{ d^{3M}X} {(L_xL_yL_z)^{lM}} [G(X/L)]^l=\n=
(2\pi g_l)^{3M(l-1)} 
\int d^{3M}K \frac {[w(K_1,...,K_M)]^l}  {(L_xL_yL_z)^{(l-1)M}} \label{38}
\ea 
with
\ba
(g_l)^{3M(l-1)}=\int d^{3M}u [G(u)]^l      \label{39}
\ea

The constant $g_l$ depends on the shape of particle distribution in
configuration space. For example, we obtain
$(g_l)^{-1}=\sqrt{2\pi}l^{1/[2(l-1)]}$ for Gaussians and
$(g_l)^{-1}=2\sqrt{3}$ for a rectangular box. In the following we shall
assume that $G(u)$ is a Gaussian\footnote{This restriction can be
avoided at the cost of some complications of the algebra. Since,
however, the exact shape of the particle emission region is not well
determined and since, moreover, (\ref{rr2}) is not in obvious
disagreement with the data from quantum interference, we shall stick to
it.}
\ba
G(u)=\frac1{(2\pi) ^{3M/2} } e^{-\sum_{m=1}^M\sum_{\alpha=x,y,z}[u_{m\alpha}]^2/2}
  .   \label{rr2}
\ea
Introducing (\ref{rr2}) into (\ref{38}) we obtain 
\ba
\hat{C}(l)=\frac{(\sqrt{2\pi} )^{3M(l-1)}}{l^{3M/2}} \int d^{3M}K 
\frac {[w(K_1,...,K_M)]^l}  {(L_xL_yL_z)^{(l-1)M}}. \label{38u}
\ea 

\section {Measurement of the moments of the Wigner function}

In this section we explain how one can estimate the moments of the Wigner
function (\ref{e1}) by counting the coincidences (\ref{31}) between the
measured events. The argument is a short summary of the results obtained
in \cite{bck}, where the relation between $\hat{C}(l)$ and $C^{exp}(l)$
was studied assuming the Wigner function\footnote{In \cite{bck} the form
(\ref{36a}) was assumed for the classical phase-space distribution. As we
have explained in the previous section, to discuss the correct
quantum-mechanical description of the problem, the Wigner function must
be used instead. This does not invalidate the results of \cite {bck}
because they are  independent of the classical or quantum nature of 
$W(X,K)$.} in the form given by (\ref{36a}).

The major problem in the analysis of coincidences between the measured
events is that these events are described by particle momenta which are
continuous variables. Therefore, the definition (\ref{31}) is not
directly applicable: a binning is necessary. Once events are
discretized, the identical ones can be defined as those which have the
same population of the predefined bins. Thus counting of coincidences
becomes straightforward\footnote{A detailed description of this
procedure was given in \cite{dl} and applied in \cite{kit}.}. The
number of identical events obviously depends on the binning, however,
so the procedure is ambiguous \cite{bc,bc2,bcw}. To obtain a viable
estimate of $\hat{C}(l)$, we  have to select the binning in such a
way that the result of (\ref{31}) is as close as possible to that given
by (\ref{38u}).

In \cite{bck} the optimal binning was determined and
the corresponding relation between $C^{exp}(l)$ and $\hat{C}(l)$ was
derived. 

Here we only quote the result in the simplest (but most importantant in
practice) case when each component of momentum is split into bins of
equal size (not necessarily the same for each component). Under this
condition the optimal size of the (3-dimensional) bin is given by

\begin{equation}
  \omega = \Delta_x\Delta_y\Delta_z=
\frac{(2\pi g_l)^3}{(L_xL_yL_z)}.   \label{omega}
\end{equation}
With this choice of the binning, the relation between 
$C^{exp}(l)$ and $\hat{C}(l)$ is
\begin{eqnarray}
\hat{C}(l) =
C^{exp}(l)\frac{\sum_{bins} <[w(\vec{K})]^l>}
{\sum_{bins}<w(\vec{K})>^l}. \label{rel}
\end{eqnarray}

Equations (\ref{omega}) and (\ref{rel}) define  the method of
estimating the moments of Wigner function $\hat{C}(l)$ from the
observed coincidence probabilities $C^{exp}$. One sees that, as
discussed in detail in \cite{bck}, the accuracy of the method improves
for large volume $(L_xL_yL_z)$ of the system.

It is interesting to observe that the condition (\ref{omega}) for
optimal binning (i.e. for (\ref{rel}) to be valid) involves only the
product $\Delta_x\Delta_y\Delta_z$. This may be employed to improve the
accuracy of the method by selecting small bins in the directions where
the momentum dependence is strong and large bins in the directions whe
this dependence is weak\footnote{E.g., in case of cylindrical symmetry
there is no need to split the azimuthal angle.}. This will bring the
correction factor in (\ref{rel}) closer to 1.

\section {Moments of Wigner function and  Renyi entropies}

In this section, using again the general form (\ref{36a}) of the
Wigner function, we discuss the coincidence
probabilities $C(l)$ defined in terms of the density matrix by (\ref{i2}).
To this end we observe that, as seen from (\ref{e2}), the density matrix
for $M$ particles can be expressed as a Fourier transform of Wigner
function \cite{wig}:

\ba
\rho_M(p;p')\equiv \rho(p_1,...,p_M;p_1',...,p_M')=
\int dX e^{i q X} W(X,K)\equiv\n\equiv 
\int d^3X_1...d^3X_M e^{i[q_1X_1+...+q_MX_M]}
W(X_1,...,X_M,K_1,...K_M)= \n=e^{-v(K_1,...,K_M)}
e^{-\frac12 \sum_{m=1}^M \sum_\alpha
L^2_\alpha q_{m\alpha}^2 + i \sum_{m=1}^M \sum_\alpha q_{m\alpha}\bar{X}_{m\alpha}(K) },
 \label{r1}
\ea
where $q=p-p'$ and $K=(p+p')/2$, and where we have explicitely used the
Gaussian form (\ref{rr2}) of  $G(u)$.

When this is introduced into (\ref{i2}) we have
\ba
C(l)=\int  d^3K_1...d^3K_M  \Omega(K_1,...,K_M;l), \label{yrr1}
\ea 
where
\ba
\Omega(K;l)=\int
\prod_{i=1}^l\left[ d^3q_1^{(i)}...d^3q_M^{(i)}\right]
\delta^3(q^{(1)}_1+...+q^{(l)}_1)...\delta^3(q^{(1)}_M+...+q^{(l)}_M)\n
e^{-\sum_{i=1}^lv(K_1^{(i)},...,K_M^{(i)})}
 e^{\sum_{m=1}^M\sum_{i=1}^l\sum_\alpha \left\{i q_{m\alpha}^{(i)}
\bar{X}_{m\alpha}^{(i)}-[q_{m\alpha}^{(i)}]^2L_\alpha^2/2\right\}}.  \label{x3}
\ea
and where we have changed the  variables from 
$p^{(1)}_m,...,p^{(l)}_m$ to
\ba
K_m= \frac 1{l} \sum_{i=1}^l p_m^{(i)} \;;\;\;\;
q^{(i)}_m=p^{(i)}_m-p^{(i+1)}_m
\;;\;\;\; K_m^{(i)}=\frac{p_m^{(i)}+p_m^{(i+1)}}2,   \label{wwx}
\ea 
with $p_m^{(l+1)}=p_m^{(1)}$.

One sees from this formula that in the limit of large $L$ only 
the region $q\approx
0$ contributes significantly to the integral. This justifies an expansion
of $v(K_1^{(i)},...,K_M^{(i)})$ in the exponent. We write
\ba
K_m^{(i)}=K_m+ k_m^{(i)},   \label{x3s}
\ea
where $k_m^{(i)}$ are linear combinations of $q_m^{(j)}$. This gives up
to  second order
\ba
v(K_1^{(i)},...,K_M^{(i)})=v(K_1,...,K_M)+ V_{m\alpha} k_{m\alpha}^{(i)}+
\frac12 V_{m\alpha,n\beta} k_{m\alpha}^{(i)}k_{n\beta}^{(i)},  \label{expan}
\ea
where 
\ba
V_{m\alpha}=\partial_{m\alpha} v(K_1,...,K_M)\;;\;\;
V_{m\alpha,n\beta}=\partial_{m\alpha}\partial_{n\beta}  v(K_1,...,K_M).\label{vdef}
\ea
The indices $m,n=1,...,M$ denote particles, $\alpha,\beta= x,y,z$ denote 
directions.

Introducing (\ref{expan}) into (\ref{x3}) we obtain a gaussian integral
which can be explicitely evaluated. The result was derived in
\cite{bkint} and reads
\ba
\Omega(K_1,...,K_M;l)=
\frac{(2\pi)^{3M(l-1)/2}}{l^{3M/2}} 
\frac{\left[w(K_1,...K_M)\right]^l}{(L_xL_yL_z)^{3M(l-1)}}
\left\{Det\left[1+\sum_{s=1}a_s T^s\right]\right\}^{-1} \label{w2}
\ea
where
\ba
a_s= \frac1{2^{2s}}\frac{(l-1)!}{(2s+1)!(l-2s-1)!}  \label{w33}
\ea
and $T$ is the $3M \times 3M$ matrix
\ba
T_{m\alpha,n\beta}=\frac1{L_\alpha} V_{m\alpha,n\beta}\frac1{L_\beta} . 
\ea
Note that the sum over $s$ is finite, because all $a_s$ vanish for 
$2s > l-1$.

If the eigenvalues $t^2_{m\alpha}$ of the matrix $T$  are known, the determinant in Eq. (\ref{w2}) can be explicitely evaluated and one obtains
\ba
\Omega(K_1,...,K_M;l)=
\frac{(2\pi)^{3M(l-1)/2}l^{3M/2} 
\left[w(K_1,...K_M)\right]^l
\prod_{m=1}^M\prod_\alpha t_{m\alpha}}
{(L_xL_yL_z)^{3M(l-1)}\prod_{m=1}^M\prod_\alpha
 \left[(1+\frac12t_{m\alpha})^l-(1+\frac12t_{m\alpha})^l\right]}.
 \label{w2x}
\ea

\section {Finite volume corrections}

The first thing we observe is that in the limit $L\rightarrow \infty$ 
one obtains simply
\ba
\Omega(K_1,...,K_M)=
\frac{(2\pi)^{3M(l-1)/2}}{l^{3M/2}} 
\frac{\left[w(K_1,...K_M)\right]^l}{(L_xL_yL_z)^{3M(l-1)}}  \label{w3}
\ea
and thus, comparing with (\ref{38}), we have in this limit
\ba
C(l)=\hat{C}(l)    ,
\ea
as expected.

One also sees from (\ref{w2}) that the next order correction depends
explicitely (through the matrix $T$) on the shape of the momentum
distribution and, therefore, it cannot be evaluated in full generality. 

For illustration, we have worked out two examples.

The simplest case when the particles are uncorrelated and the
 momentum distribution is a Gaussian,  
\ba
v=\sum_{m=1}^M \sum_\alpha [K_{m\alpha}]^2/(2\mu_\alpha^2)+
\frac12\log[2\pi \mu_\alpha^2]
  \label{w4}
\ea 
gives
\ba
T_{m\alpha,n\beta}=\delta_{mn}\delta_{\alpha\beta}\frac1{L_\alpha^2\mu_\alpha^2}
\ea
and thus
\ba
Det\left[1+\sum_{s=1}a_s T^s\right]=\prod_\alpha \left\{\sum_{s=0} 
 \frac{a_s}{ (L_\alpha^2\mu_\alpha^2)^s}\right\}^M.
\ea
Consequently, we have
\ba
C_l=\hat{C}_l \prod_\alpha \left\{\sum_{s=0} 
 \frac{a_s}{ (L_\alpha^2\mu_\alpha^2)^s}\right\}^{-M}=
\hat{C}_l \prod_\alpha \left\{\left[1+\sum_{s=1} 
 \frac{a_s}{ (L_\alpha^2\mu_\alpha^2)^s}\right]\right\}^{-M}  \label{w12}
\ea
and one sees that the corrections vanish in the limit $L_\alpha^2\mu_\alpha^2\;
\rightarrow \; \infty$.

As the second example we have studied corrections for $l=3$ in a more
realistic case of an uncorrelated (factorizable) cylindrically symmetric
boost-invariant distribution with Boltzmann shape in transverse momentum
\ba
v(K_1,...,K_M)=\sum_{m=1}^M v(K_m)      \;;\;\;\;\n
v(K)=\frac{\sqrt{K_\perp^2+m^2}}{T}+\log E+\log [2\pi T(m+T)e^{-m/T}]
 \label{1w4}
\ea
where $T$ is a parameter, $E$ is the energy of the particle and $m$
its mass. The constant is added to guarantee the
correct normalization in the Y interval of unit length. 
Using (\ref{1w4}), the matrix $V_{\alpha\beta}$ and the determinant
$Det[1+T/12]$, necessary to evaluate (\ref{w2}) for $l=3$,
can be found \cite{bkint}. 

  \begin{figure}[htb]
\centerline{%
\epsfig{file=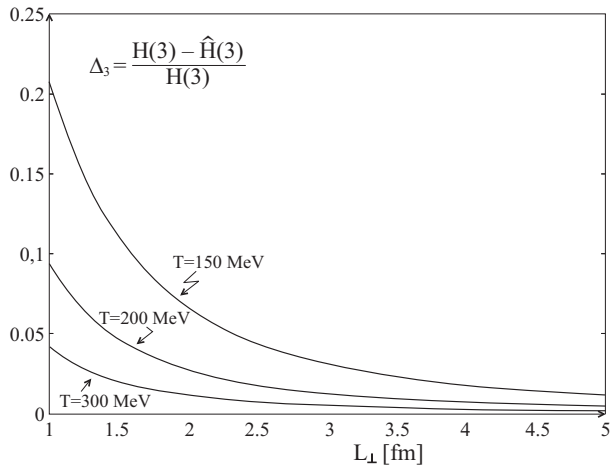,width=8cm}}
\caption{The relative difference
$\Delta_3=1-\hat{H}(3)/H(3)$ plotted versus 
$L_\perp$ for several values of the "temperature" $T$.
\label{fig1}}
\end{figure}

In Figure 1, the relative difference
\ba
\Delta_3 \equiv \frac{H(3)-\hat{H}(3)}{H(3)} =1-\frac{\log\hat{C}(3)}
{\log C(3)}
\ea
is plotted versus $L_\perp$ for various values of $T$. 
 One sees that for $T\geq 150 $ MeV (corresponding to the average
transverse momentum larger than 300 MeV), and $L_\perp \geq 3$ fm
(appropriate for heavy ion collisions) $\Delta_3$ is indeed very
small. We thus conclude that the moments of Wigner function
reproduce  the  Renyi entropies of a
multiparticle system created in high-energy nuclear collisions
 with  rather good accuracy.

One sees also from Fig. 1 that the corrections increase at smaller
values of $L_\perp $. At $L_\perp\approx 1$ fm (appropriate for
elementary collisions) they reach about 20\% for $T=150$ MeV and fall
quickly at larger $T$. Thus, although for hadron-hadron and
lepton-hadron collisions our estimates of Renyi entropy clearly require
a more precise determination of the size of the system, they seem
nevertheless to be within reach of the present experiments.

These two simple examples give only a general idea how to control the
corrections. Estimates of corrections in more complicated situations are
of course posssible, e.g., by Monte Carlo simulations.

\section{Discussion and conclusions}

 Several comments are in order. 

(i) Although we have discussed the general case of arbitrary $l$, it
should be realized that, in practice, one may at best hope for the
determination of the lowest order Renyi entropies $l=2,3$, perhaps also
$l=4$. This implies that an extrapolation to $l=1$, giving the Shannon
entropy (\ref{i3}), may require an independent input to be reliable
\cite{zy}.

(ii) One sees from (\ref{w33}) that for $l=2$ $a_1=0$, i.e.,
$C(2)=\hat{C}(2)$. Thus the Renyi entropy $H(2)$ does not suffer from
the corrections discussed in this paper. 

(iii) As is seen from the discussion in section 4, the difference
between Renyi entropies and moments of the Wigner function depends
primarily on the size of the system in configuration space. Also
accuracy of the measurements of the moments of Wigner function depends
on this size \cite{bck}, as disussed in Section 3. One concludes that
information from the HBT measurements \cite{dan,sa} is important for a
successful application of the method.

(iv) One should keep in mind that we are discussing here the phase-space
distribution and the Wigner function averaged over time. If the freeze-out
takes relatively long time, the effective size of the system may be
much larger than naively expected. On one hand this would improve the
accuracy of the method. On the other hand it makes the estimate of the
volume much more difficult, as the standard interpretation of the HBT
measurements may be not adequate \cite{bk}.

(v) It should be emphasized that the estimate of entropy investigated in
the present paper takes explicitely into account the correlations
between the particles observed in the experiment. This may be contrasted
with the estimate obtained in \cite{palb} where the entropy of the
multiparticle system is estimated from the single-particle inclusive
distribution, thus ignoring the correlations between particles (except
those induced by quantum interference). It would be very interesting to
compare the results obtained from these two methods. This may provide an
insight into the role of multiparticle correlations in counting the
number of effective degrees of freedom of a multiparticle system.

(vi) The entropies we discuss refer to the particles actually measured
in experiment in question. Therefore the result does depend on the
nature of the detector. E.g., the results change when entropy is
determined from all produced particles instead of only from the charged
ones. This should not be surprising: the effective number of degrees of
freedom naturally depends on the number and nature of the particles
considered. Actually, investigations of the dependence of entropy on the
number of particles may provide some interesting hints on the structure
of produced final states.

(vii) Through this paper we have only discussed entropies of the
M-particle distribution at  fixed $M$. One is often interested in
entropies summed over all multiplicities. They may be obtained from
coincidence probabilities referred to all multiplicities, constructed
following the formula 
\ba 
C(l)= \sum_M [P(M)]^l C_M(l), \label{mult} 
\ea 
where $P(M)$ is the multiplicity distribution and where, for the sake of
clarity, we have added a subspript $M$ to denote the coincidence
probability  at fixed $M$. One sees that for large $l$ only
multiplicities close to the most probable one contribute to the sum.

In conclusion, we have analyzed the relation between the moments of
Wigner function (\ref{e1}) (which can be measured by counting the number
of identical events \cite{bck}) and the coincidence probabilities
(\ref{i2}) (which define the Renyi entropies). It was shown that, for a
large class of models, these moments are identical to the coincidence
probabilities in the limit of an infinite volume of the system. The
finite volume corrections were discussed. They were shown to fall as the
inverse square of the linear size of the system at freeze-out and turn
out to be negligible for systems encountered in relativistic heavy ion
collisions.

\vspace{0.5cm}

\vspace{0.3cm}
{\bf Acknowledgements}
\vspace{0.3cm}

Discussions with Robi Peschanski and Jacek Wosiek were very useful and
are highly appreciated. This investigation was partly supported by the
MEiN research grant 1 P03B 045 29 (2005-2008).


\begin{thebibliography} {99}
\bibitem{bc}
A.Bialas and W.Czyz, Phys. Rev. D61 (2000) 074021.
\bibitem{ren}
A.Renyi, Proc. 4-th Berkeley Symp. Math. Stat. Prob. 1960, Vol.1, Univ.
of California Press, Berkeley-Los Ageles 1961, p.547.
\bibitem{beck}
C.Beck and F.Schloegl, Thermodynamics of chaotic systems, Cambridge U.
Press, Cambridge (1993).
\bibitem{plasma}
For a recent discussion see, e.g. B.Muller and K.Rajagopal,
hep-ph/0502174 and references therein. 
\bibitem{ber1}
G.F.Bertsch, Phys. Rev. Letters 72 (1994) 2349; 77 (1996) 789 (E).
\bibitem{bpb}
D.A.Brown, S.Y.Panitkin and G.F.Bertsch, Phys. Rev. C62 (2000) 014904.
\bibitem{wig} For a discussion of the physical meaning of the Wigner function
 see, e.g., M.Hillery, R.F.O'Connell, M.O.Scully and
E.P.Wigner, Phys.Rept. 106 (1984) 121 and references therein.
\bibitem{bck}
A.Bialas, W.Czyz and K.Zalewski, Acta Phys. Pol. B36 (2005) 3109; 
 hep-ph/0506233, to be published in Phys. Lett. B.
\bibitem{dl}
A.Bialas and W.Czyz, Acta Phys. Pol. B31 (2000) 687.
\bibitem{bc2}
A.Bialas and W.Czyz, Acta Phys. Pol.  B31 (2000) 2803; 
B34 (2003) 3363.
\bibitem{ma}
S.K.Ma, Statistical Mechanics, World Scientific, Singapore 1985; S.K.Ma,
J. Stat. Phys. 26 (1981) 221.
\bibitem{bcw}
A.Bialas, W.Czyz and J.Wosiek, Acta Phys. Pol. B30 (1999) 107.
\bibitem{mod}
It is obviously valid for models which assume thermal equilibrium.
It is also valid in the blast-wave models, provided $\bar{X_i} = 
\bar{X_i}(K_i)$. The Hubble-like expansion is obtained
 for $\bar{X_i}\sim K_i$. 
For a review of models see, e.g.,
U.A. Wiedemann and U. Heinz, Phys. Rep. 319(1999)145;
U. Heinz and B. Jacak, Ann. Rev. Nucl.Part.Sci. 49(1999)529;
R.M. Weiner, Phys. Rep. 327(2000)250; T.Csorgo, H.I.Phys. 15 (2002)1.
\bibitem{kit}
 K.Fialkowski and R.Wit, Phys.Rev. D62 (2000) 114016;
 NA22 coll, M. Atayan et al., Acta Phys. Pol. B36 (2005) 2969.
\bibitem{zy} K.Zyczkowski, Open Sys. and Information Dyn. 10 (2003) 297.
\bibitem{bkint}
A.Bialas and K.Zalewski, hep-ph 0512248, to be published in 
Acta Phys. Pol. B.
\bibitem{dan}
D.A.Brown and P.Danielewicz, Phys. Lett. B398 (1997) 252; Phys. Rev D58
 (1998) 094003; S.Y.Panitkin and D.A.Brown, Phys. Rev C61 (1999) 021901;
G.Verde et al, Phys. Rev. C65 (2002) 054609; P.Danielewicz et al., Acta
Phys. Hung. A19 (2004) nucl-th/0407022.
\bibitem{sa}
Yu.M.Sinyukov and S.V.Akkelin, Heavy Ion Phys. 9 (1999); S.V.Akkelin and
Yu.M.Sinyukov, nucl-th/0310036.
\bibitem{bk} See, e.g., A.Bialas and K.Zalewski, Phys. Rev. D72 (2005) 036009.
\bibitem{palb} S.Pal and S.Pratt, Phys. Lett. B578 (2004) 310.




\end{thebibliography}
\end{document}